\documentclass{article}
\usepackage[utf8]{inputenc}

\usepackage[numbers,sort&compress]{natbib}
\usepackage{graphicx}

\usepackage{amssymb}
\usepackage{amsmath}
\usepackage{hyperref}
\usepackage[a4paper,top=1.5cm,bottom=1.5cm,left=2cm,right=2cm,marginparwidth=1.75cm]{geometry}
\usepackage{subfig}
\usepackage[utf8]{inputenc}

\begin{document}

\begin{titlepage}

\phantom{.}
\vspace{3cm}

{\bf\boldmath\Large
\begin{center}
  Summary of the 2018 CKM working group on semileptonic and leptonic $b$-hadron decays 
\end{center}
}

\vspace{1cm}

\begin{center}
C.~Bouchard$^1$, L.~Cao$^2$, P.~Owen$^3$ \bigskip\\
{\it\footnotesize 
$ ^1$School of Physics and Astronomy, University of Glasgow, Glasgow, United Kingdom \\ \vspace{0.2cm}
$ ^2$Karlsruher Institute of Technology, 76131 Karlsruhe, Germany \\ \vspace{0.2cm}
$ ^3$Universit\"at Z\"urich, Z\"urich, Switzerland  \\ \vspace{0.2cm}
}

\vspace{0.5cm}

\today

\end{center}

\vspace{2cm}

\begin{abstract}
  \noindent
  A summary of WG II of the CKM 2018 conference on semileptonic and leptonic $b$-hadron decays is presented. This includes discussions on the CKM matrix element magitudes $|V_{ub}|$ and $|V_{cb}|$, lepton universality tests such as $R(D^{*})$ and leptonic decays. As is usual for semileptonic and leptonic decays, much discussion is devoted towards the interplay between theoretical QCD calculations and the experimental measurements.
  \end{abstract}

\vspace{\fill}

\end{titlepage}

\section{Introduction}

Semileptonic and leptonic $b$-hadron decays describe the decay of any beauty hadron which results in at least one neutrino in the final state. The presence of the neutrino in the final state makes such decays theoretically appealing, as it is an unambiguous signal of a short-distance interaction. This means that branching fractions can be precisely predicted, which allows the crucial CKM parameters $|V_{ub}|$ and $|V_{cb}|$ to be measured. This does, however, complicate the experimental measurements as background is more easily identified with the partially reconstructed semileptonic signals.

Therefore, the huge potential of semileptonic $b$-hadron decays can only be realised if state-of-the-art theoretical and experimental techniques are employed. The interplay between the two to minimise uncertainties is crucially important. This results in semileptonic sessions at CKM being a highly engaging and thoroughly useful exercise.

The 2018 CKM session on semileptonic and leptonic decays can be split fairly equally into three parts. The first is related to the determination of the CKM element magnitudes $|V_{cb}|$ and $|V_{ub}|$ from both exclusive and inclusive decays. The second is tests of lepton universality in semitauonic decays such as $B\to D^{*} \tau\nu_{\tau}$ and the final is fully leptonic decays such as $B\to\mu\nu$. All three of these areas had new experimental and/or theoretical progress since the last CKM conference.

\section{Determining $|V_{cb}|$ and $|V_{ub}|$}
\begin{figure}[h]
	\centering
	         \subfloat{
		 \includegraphics[width=0.5\linewidth]{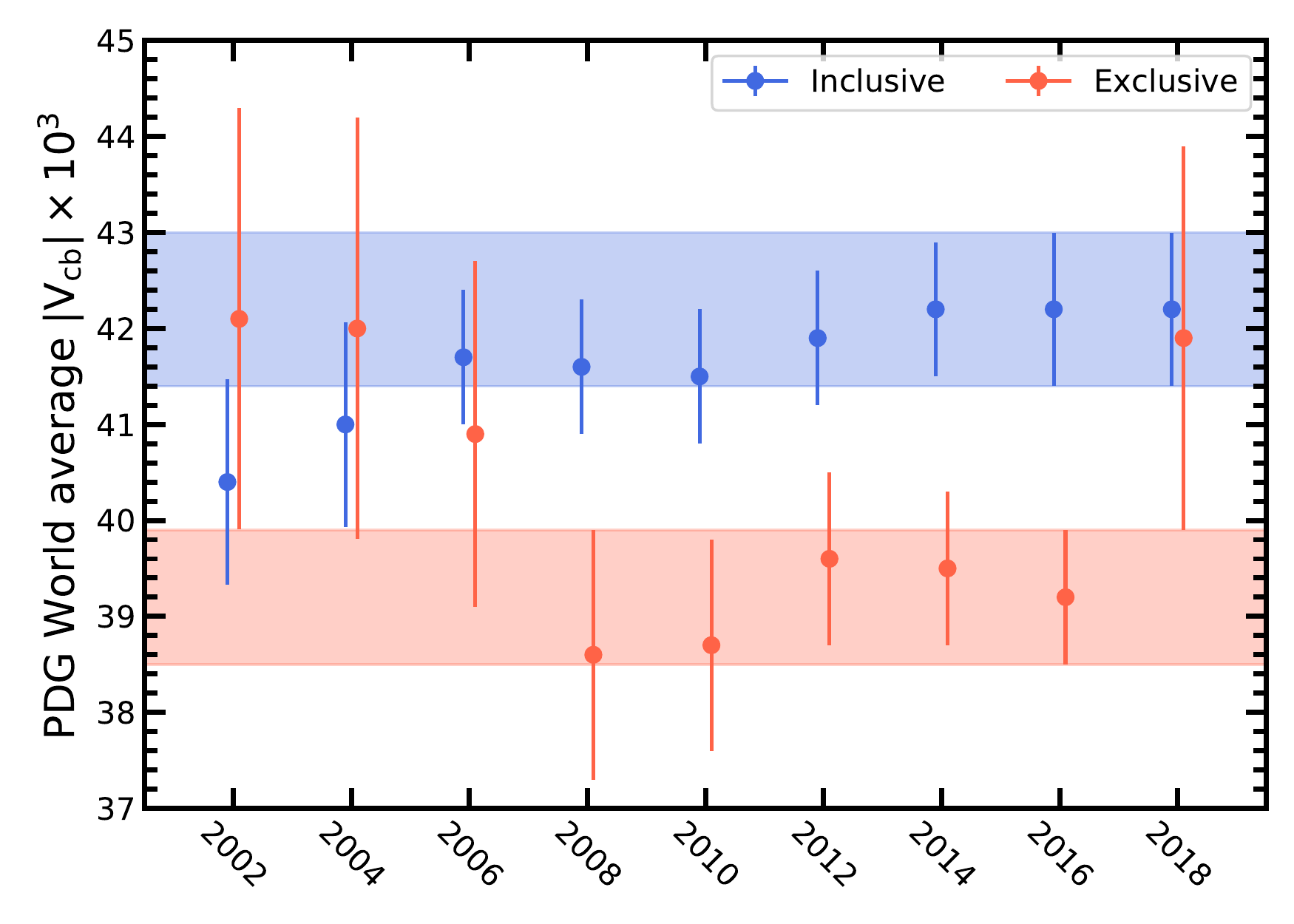}
		 }
		 \subfloat{
		  \includegraphics[width=0.5\linewidth]{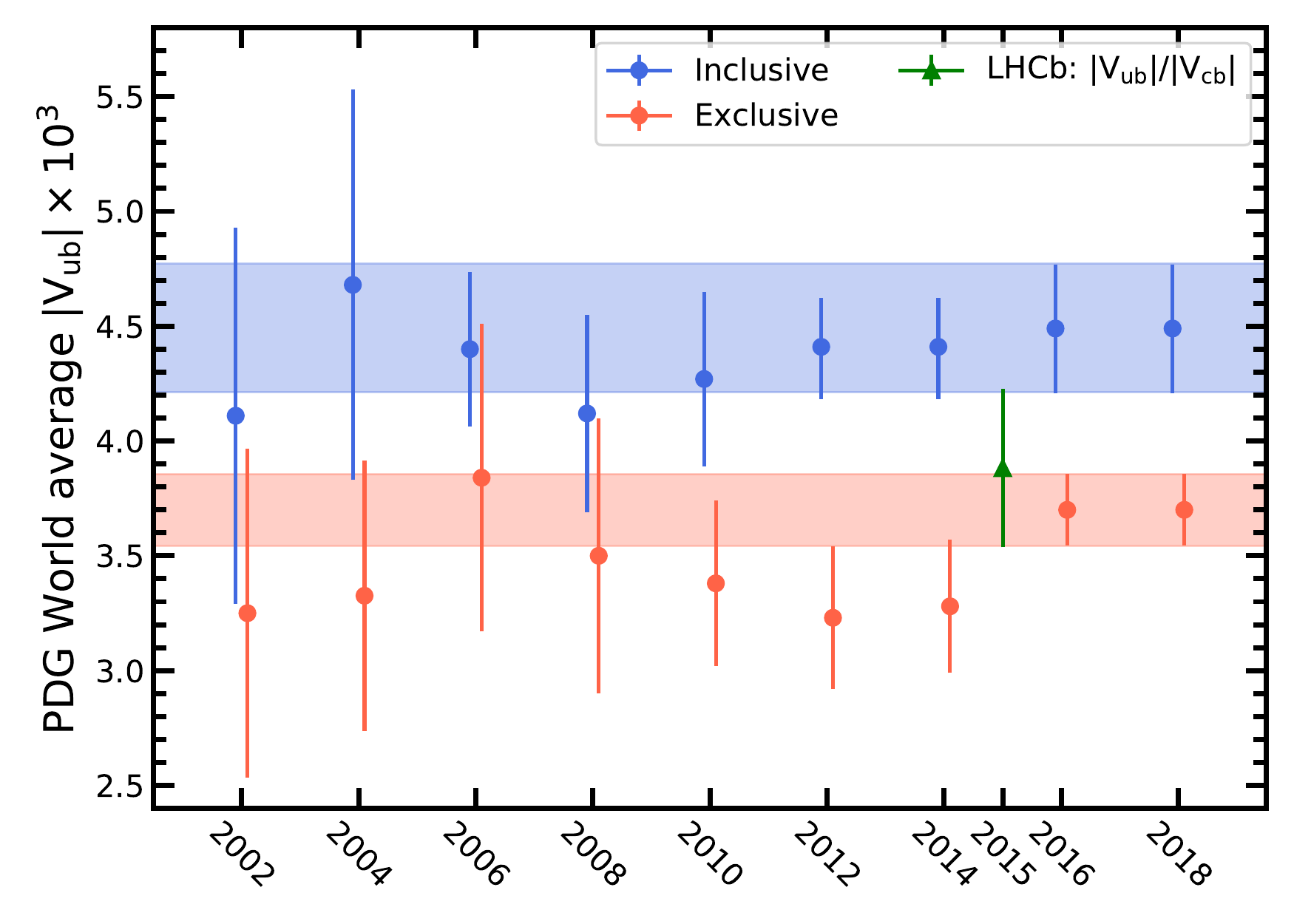}
		  }
	\caption{\label{fig:vxb}Summary of PDG average of $|V_{cb}|$ (left) and $|V_{ub}|$ (right) separating the existing results between exclusive and inclusive measurements. The green point is determined by the ratio $|V_{ub}/V_{cb}|$ from LHCb measurement \cite{Aaij:2015bfa} and the latest average $|V_{cb}|$.}
\end{figure}

As the fundamental CKM elements, $V_{cb}$ and $V_{ub}$ play an important role in testing unitarity. 
Fig.~\ref{fig:vxb} shows the PDG averaged $|V_{ub}|$ and $|V_{cb}|$ from inclusive and exclusive methods respectively. 
In this field, a long-standing puzzle is the discrepancy between exclusive and inclusive determinations with $B$ meson semileptonic decays. 
The tension for $|V_{cb}|$ is $3\sigma$ excluding the latest exclusive contribution \cite{Abdesselam:2017kjf}, and $3.5\sigma$ for $|V_{ub}|$ determinations. 

\subsection{Exclusive decays}
\begin{figure}[tb]
  %\vspace{-0.5in}
  \centering
  \includegraphics[width=0.8\textwidth]{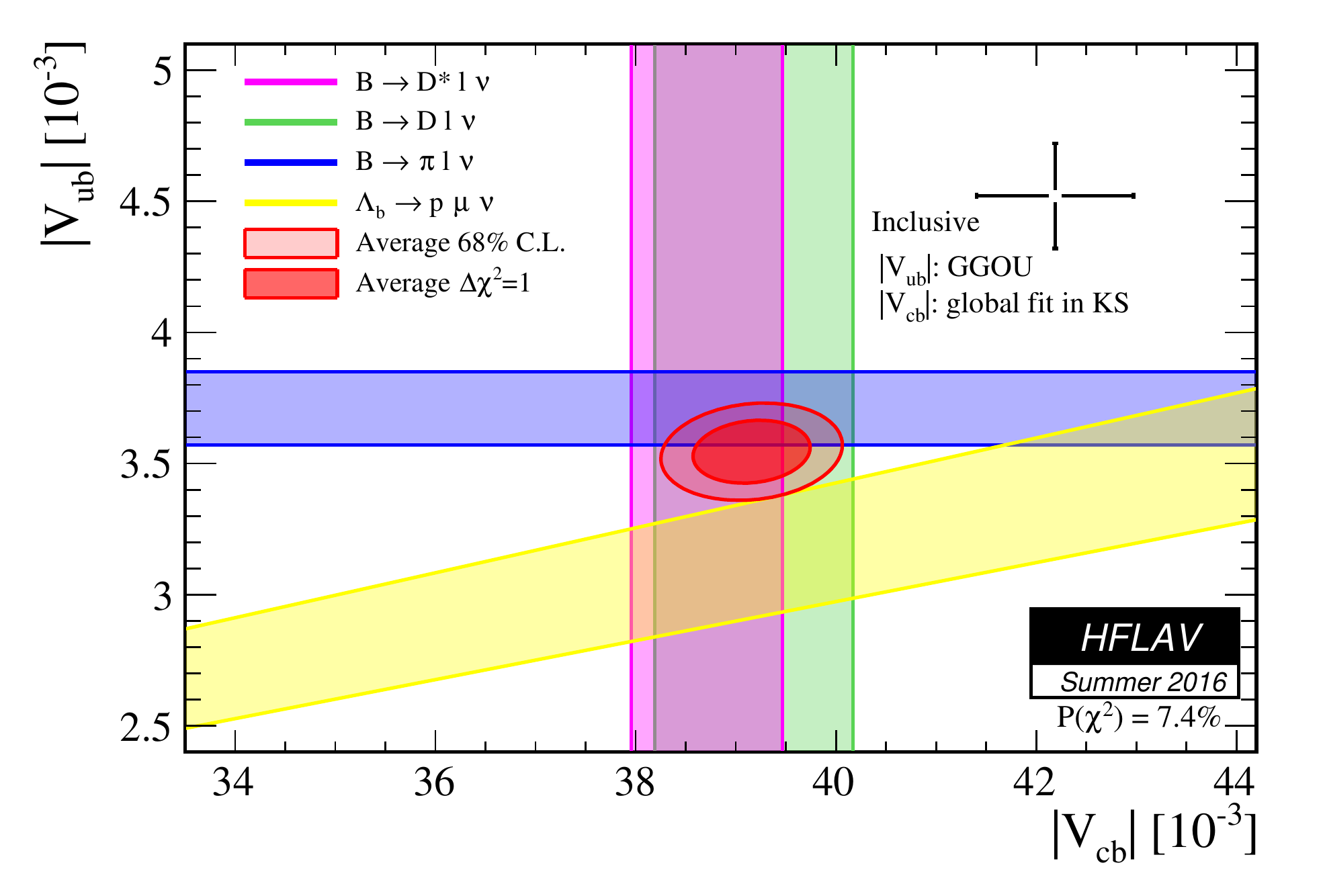}
  \caption{Comparison of various exclusive determinations of $|V_{ub}|$ and $|V_{cb}|$~\cite{HFLAV16}.}
  \label{fig:HFLAV}
\end{figure}
Recent analyses demonstrate that the tension between inclusive and exclusive determinations of $V_{cb}$ can be significantly reduced by using a sufficiently general form factor parameterization to describe the kinematic dependence in $B\to D^{(*)}l\nu$~\cite{PhysRevD.93.032006, Bigi:2017njr, Bigi:2016mdz, Grinstein:2017nlq, Bernlochner:2017jka, Abdesselam:2018nnh}.
To clarify the $V_{cb}$ puzzle, it will be important in future combinations of theory and experiment to use unbinned experimental data and to carefully scritinize the kinematic parameterization of the form factors.  
This will ensure potential kinematic model-dependence is either removed or is adequately accounted for in error budgets while allowing us to leverage available shape information.

Recent and ongoing lattice QCD calculations for $b\to c$ decays, some of which include an extended region of $q^2$, include
FNAL-MILC for $B\to D^*l\nu$~\cite{Aviles-Casco:2019vin}, 
JLQCD for $B\to D^{(*)}l\nu$~\cite{Kaneko:2018mcr},
HPQCD for $B_{(s)}\to D^{(*)}_{(s)}l\nu$~\cite{McLean:2017kdq,  Monahan:2017uby, Harrison:2017fmw},
RBC/UKQCD for $B_s\to D_sl\nu$, 
and LANL/SWME for $B\to D^*l\nu$~\cite{Bailey:2017xjk}.
These calculations utilize an effective field theory treatment of the $b$ quark, leading to an often leading source of systematic uncertainty from matching to the effective theory.
In an important step toward reducing uncertainties, HPQCD is calculating $B_s\to D_s^*l\nu$ using a fully relativistic treatment of the $b$ quark~\cite{McLean:2019jll}, free from uncertainty associated with matching to the effective theory.

Recent or ongoing lattice QCD calculations for $b\to u$ decays include FNAL-MILC for $B\to \pi l\nu$ and $B_s\to Kl\nu$~\cite{Gelzer:2017edb, Lattice:2017vqf},
RBC/UKQCD for $B_s\to Kl\nu$,
JLQCD for $B\to \pi l\nu$~\cite{Colquhoun:2018kwj},
and HPQCD for $B\to\pi l\nu$.
The added computational cost has, thus far, prevented the use of fully relativistic $b$ quarks in $b\to u/d$ transitions.
It will be important to reduce or remove the matching uncertainties by, e.g., using a fully relativistic treatment of the $b$ quark.
Using Light cone sum rules, Khodjimirian and Rusov~\cite{Khodjamirian:2017fxg} recently calculated the form factors for $B_s\to Kl\nu$ and Khodjimirian reported on the potenitial use of $B$ meson distribution amplitudes (see, e.g., Refs.~\cite{Khodjamirian:2010vf, Braun:2017liq}) to calculate $B\to\pi\pi$ form factors, a nonresonant background for $B\to\rho l\nu$.

The correlated ratio $|V_{ub}/V_{cb}|$ can be extracted with reduced uncertainty via a calculation of ratios of form factors and measurement of associated branching fractions.
This has recently been done for the $\Lambda_b\to\Lambda_c\ell\nu$ and $\Lambda_b\to pl\nu$ decays with a lattice QCD calculation of the form factors by Detmold et al.~\cite{Detmold:2015aaa} and subsequent measurement of branching fractions by LHCb~\cite{PhysRevD.96.112005, Aaij:2015bfa}.
The impact of this analysis in the determination of $|V_{ub}|$ and $|V_{cb}|$ is illustrated in Fig.~\ref{fig:HFLAV}.
This is currently being done for the $B_s\to Kl\nu$ and $B_s\to D_sl\nu$ decays, with the correlated ratio of form factors calculated by HPQCD~\cite{Monahan:2018lzv} and an ongoing LHCb analysis of the branching fractions.

\subsection{Inclusive decays}
The 2018 world average values are $\left|V_{cb}\right| = ( 42.2 \pm 0.8 ) \times 10^{-3}$ and $\left|V_{ub}\right| = ( 3.94 \pm 0.36 ) \times 10^{-3}$ \cite{PDG2018}. 
The theoretical predictions of inclusive $V_{cb}$ are based on operator product expansion (OPE) with a dependence on quark masses $m_{b}$ and $m_{c}$, and other input parameters in subleading and higher order corrections. The theoretical efforts for inclusive $V_{cb}$ focus on decreasing the uncertainties in OPE calculations \cite{Gambino:2017vkx, Bazavov:2018omf} and providing more information to cross check, e.g. the inclusive structure functions from lattice QCD. \cite{Hashimoto:2017wqo}. 

Compared to $V_{cb}$, the experimental determination of $V_{ub}$ from inclusive $B\to X_{u} l\nu$ decay is more complicated due to large $B\to X_{c} l\nu$ background. The current precision of measurements is expected to be improved with the Belle II experiment with a 40 times higher luminosity of Belle~\cite{Kou:2018nap} as well as the newly implemented tagging techniques~\cite{Keck:2018lcd}. Several global data-driven fit methods have been prepared for exacting the $V_{ub}$ in a model-independent way to improve the precision by reducing model uncertainties from the $B$ meson shape function~\cite{PhysRevD.94.014031, Bernlochner:2013gla}.

\section{Tests of lepton flavour universality}
The lepton flavor universality implying couplings of electroweak bosons to different leptons are independent of their flavor, and the only difference arises from the masses of leptons. Observing the violations of this SM assumption will bring a hint for new physics. There are three areas of recent interest:
\begin{itemize}
\item $e-\mu$ universality in $b\to cl\nu$ 

The recent experimental result on $\mathcal { R }_{e/\mu} = \mathcal { B }(B\to D^{(*)} e \nu_{e})/\mathcal { B }(B\to D^{(*)} \mu \nu_{\mu})$ from Belle \cite{Abdesselam:2017kjf, Abdesselam:2018nnh} are consistent with unity predicted by SM. In addition, based on the existing $B\to D^{(*)} l \nu$ analyses separating results for electrons and muons, Ref.~\cite{Jung:2018lfu} demonstrates an agreement on $R_{e/\mu}$ between $D$ and $D^{*}$ mode.

\item $\tau-\mu$ universality in $b\to cl\nu$ 

The semitauonic decays provide the test universality between heavier leptons, e.g. $\mathcal { R } ( D ^ { ( * ) } ) = \mathcal { B } ( B  \to D ^ { ( * )  } \tau  \overline { \nu } _ { \tau }) / \mathcal { B } ( B  \to D ^ { ( * ) } \mu  \overline { \nu } _ { \mu } )$.
The current global average indicates that $\mathcal { R } ( D )$ and $\mathcal { R } ( D ^ {*} )$ exceed the SM predictions by $2.3\sigma$ and $3.0\sigma$ respectively, and the combination shows deviation of $3.8\sigma$ as shown in Fig.~\ref{fig:RdRds}. The very recent measurements from LHCb \cite{Aaij:2017uff} and Belle \cite{Hirose:2017dxl} have been included in the average fit.

\begin{figure}[tb]
  \centering
  \includegraphics[width=0.8\textwidth] {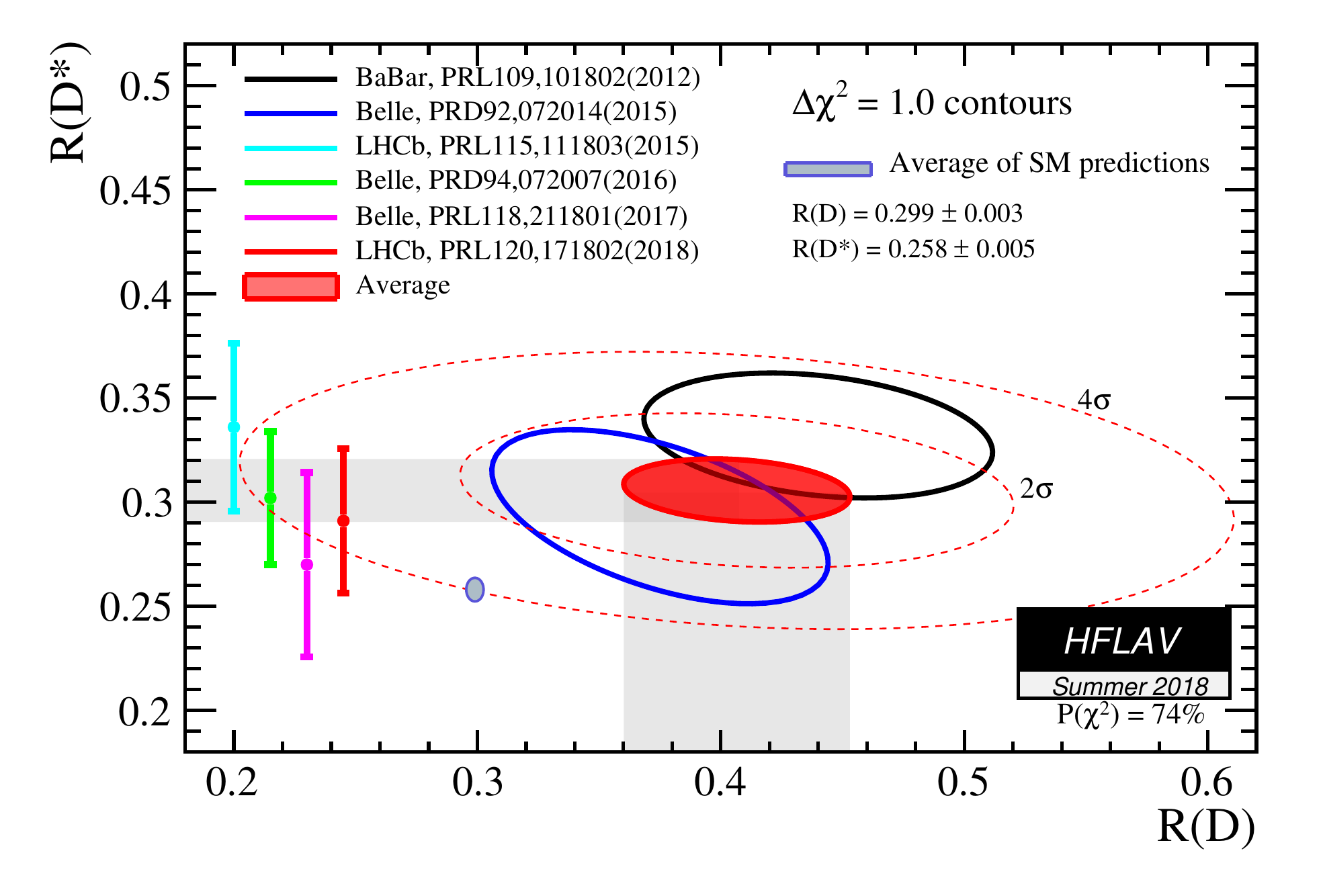}
  \caption{Experimental average fit on $\mathcal { R } ( D )$ and $\mathcal { R } ( D ^ {*} )$ and their averaged theoretical predictions assuming SM. The tension of discrepancy is shown by the red dashed lines.}
  \label{fig:RdRds}
\end{figure}

In addition, $\mathcal{R}(J/\psi) = \mathcal{B}(B_c^+\to J/\psi\tau^+\nu_\tau)$/$\mathcal{B}(B_c^+ \to J/\psi\mu^+\nu_\mu)$ has been recently measured by LHCb~\cite{Aaij:2017tyk}, which indicates $2\sigma$ deviations above the range of central values currently predicted by the SM.

\item $e-\mu$ universality in $b\to s ll$

The ratio of the branching fractions of the $B ^ { 0 } \to K ^ { * 0 } \mu ^ { + } \mu ^ { - }$ and $B ^ { 0 } \to K ^ { * 0 } e ^ { + } e ^ { - }$, namely $\mathcal { R } ( K ^ { ( * ) } )$, is a search of neutral LFU related to the flavor-changing-neutral-current mediated in $B$ decays. Recent results from LHCb \cite{Aaij:2017vbb} found around 2.2$\sigma$ and 2.4$\sigma$ deviations from unity at the low and central $q^{2}\equiv m^{2}(ll)$ bins respectively.

\end{itemize}

Above measurements imply the noticeable deviations with SM so far, which attract more and more interests and efforts on new physics models (e.g. $Z^{\prime}$, leptoquark) \cite{Jung:2018lfu, Buttazzo:2017ixm, Becirevic:2018afm}. On the other hand, the experimental tests with more statistics, more decay modes, providing new observable beyond ratios of branching fraction will be needed to discriminate between new physics scenarios. 
Long-distance QED corrections to $\mathcal{R}(D)$ from soft photons, 20--40~MeV, can be as large as a few percent~\cite{PhysRevLett.120.261804}.  It is important to ensure that such effects hare handled consistently in experiment and theory.

\section{Leptonic decays}

\subsection{Decay constants}

Fully leptonic decays of b-hadrons are in principle more theoretically appealing than semileptonic decays. Here the theoretical challenge is to calculate the decay constants of the relevant b-hadron species. These can be very preciely calculated using lattice QCD. The latest results~\cite{Bazavov:2017lyh} of which for both $B$ and $D$ mesons were presented:
\begin{align*}
    f_{D^0}  &= 211.6  (0.6)~{\rm MeV},\ 
    f_{D^+}  = 212.7  (0.6)~{\rm MeV},\  
    f_{D_s}  = 249.9  (0.5)~{\rm MeV},\\
    f_{B^+}  &= 189.4  (1.4)~{\rm MeV},\  
    \, f_{B^0}  = 190.5  (1.3)~{\rm MeV},\  
    \, f_{B_s}  = 230.7  (1.3)~{\rm MeV}.
    \label{eq:fBs}
\end{align*}
where the uncertainties shown in the parentheses are dominated by systematic effects for $D$ mesons and statistical effects in $B$ mesons. The precision of these decay constant measurements is better than the current experimental precision, and can be improved in the future. 

These decay constants are used for a verity of purposes, such as testing unitarity in the 2nd row of the CKM matrix, predicting the rare decay branching fractions $B_{(s)}^{0} \to \mu\mu$ and determining $|V_{ub}|$ from the branching fraction of the decay, $B^{+}\to \mu^{+}\nu_{\mu}$.

\subsection{$B^{+}\to \mu^{+}\nu_{\mu}$}

The decay $B^{+}\to \mu^{+}\nu_{\mu}$ is an extremely clean way of measuring the CKM element magnitude $|V_{ub}|$. The decay poses considerable experimental challenges due to helicity suppression of the decay amplitude and single track signature. 

The more sensitive search for such a decay is from the Belle collaboration~\cite{Sibidanov:2017vph}, which uses untagged events. The main challenge is to distinguish the signal signature of a high-energy isolated muon, with background. The most dangerous background is $B\to X_{u} \mu\nu_{\mu}$ decays, where the $X_{u}$ is not reconstructed. 

\begin{figure}[tb]
  \centering
  \includegraphics[width=0.8\textwidth] {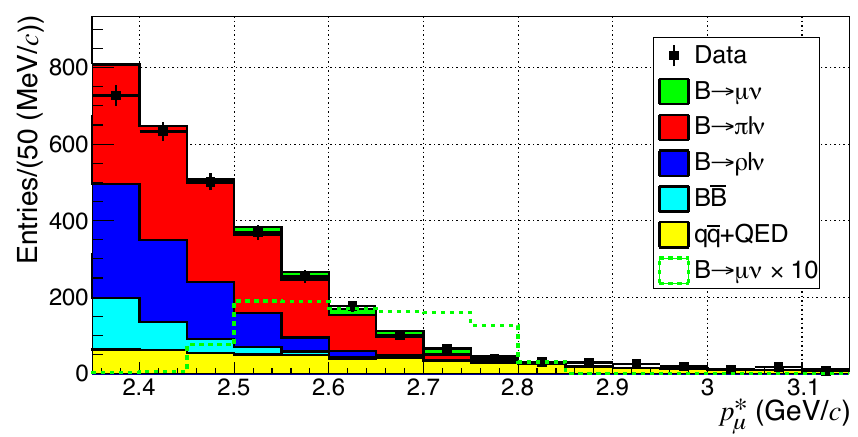}
  \caption{Projection of the fit to the muon momentum, $p_{\mu}^{*}$, for the search of $B\to\mu\nu_{\mu}$ decays.}
  \label{fig:BMuNu_fit}
\end{figure}

A fit is performed to the energy of the muon, $p_{\mu}^{*}$ and a neural network output which many kinematic variables designed to distinguish the signal and background. The projection of the fit onto $p_{\mu}^{*}$ is shown in Fig.~\ref{fig:BMuNu_fit}. A signal significance of 2.4~$\sigma$ is found, and a 90\% confidence interval of $[2.9, 10.7] \times 10^{-7}$ is calculated. This is consistent with the SM prediction of $(3.80 \pm 0.31) \times 10^{-7}$ but leaves lots of room for contributions from new physics particles which are not helicity suppressed. Further data from Belle-II is highly anticipated to improve this measurement.

\subsection{The decay $\boldsymbol {B^{+}\to \ell^{+}\nu_{\ell}\gamma}$}

The strong helicity suppression of $B^{+}\to \mu^{+}\nu_{\mu}$, can be lifted with the radiation of a photon. This is more theoretically complicated as calculations of hadronic form factors for the transition to the photon. These form factors depend on  the B-meson light-cone distribution amplitude, which is a crucial input for QCD-factorisation.

The Belle collaboration has recently searched for the decay $B^{+}\to \ell^{+}\nu_{\ell}\gamma$ with $E_{\gamma} > 1~{\rm GeV}$. The analysis uses a novel technique for the tagging of the other $B$ meson in the event~\cite{Keck:2018lcd}. The algorithm uses Boosted Decision Tree algorithms to reconstruct over 2000 decay modes. The performance with this algorithm improved by around 40\%  compared to the previous tagging method.

\begin{figure}[tb]
  \centering
  \includegraphics[width=0.9\textwidth] {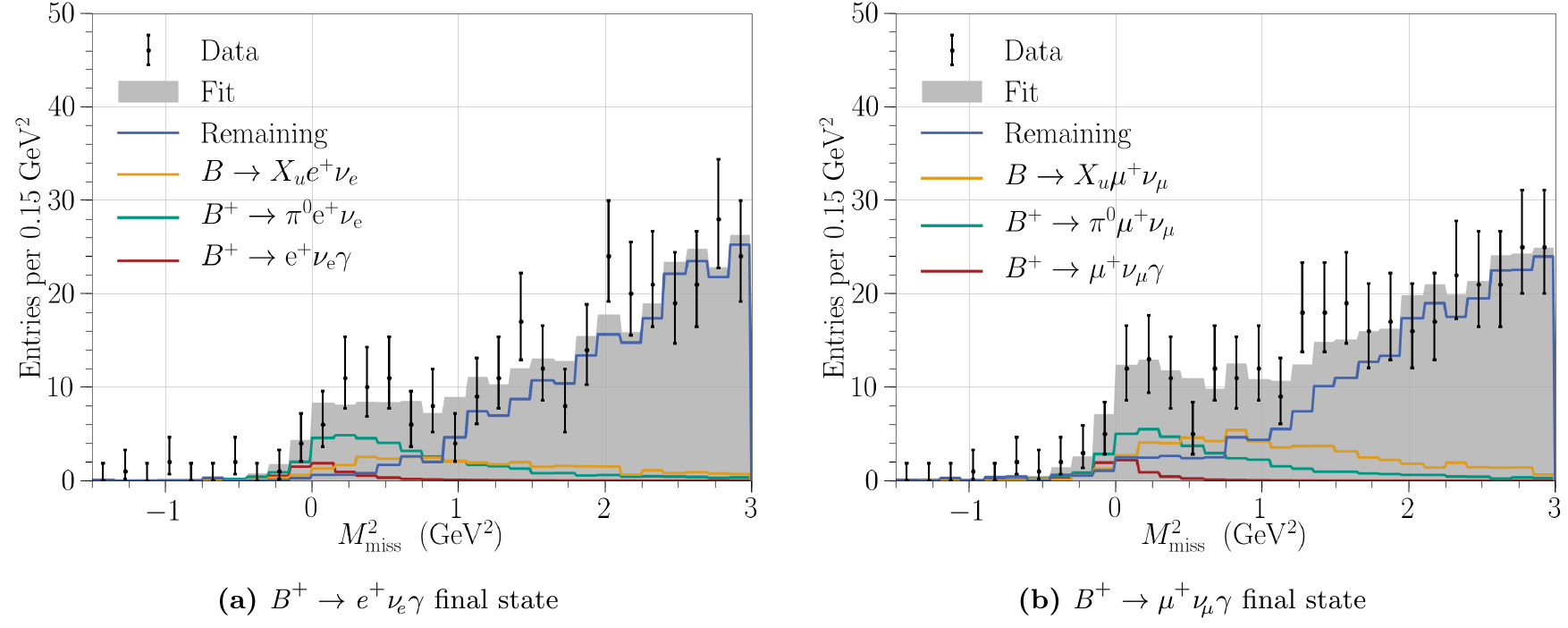}
  \caption{Projection of the fit to the missing mass, $M_{miss}$, for the determination of the $B\to\ell\nu_{\ell}\gamma$ signal yield.}
  \label{fig:BMuNuGamma_fit}
\end{figure}

The signal yield is determined from a fit to the distribution of the missing mass, as shown in Fig.~\ref{fig:BMuNuGamma_fit}. The missing mass is reconstructed by comparing the kinematics of the signal $B$ decay, as determined from the tagging method, with that of the signal lepton and photon. As with the $B^{+}\to \mu^{+}\nu_{\mu}$ analysis, a small signal excess of 1.4\,$\sigma$ is observed. This results in a limit being set of $\Delta\mathcal{B}(B^{+}\to \ell^{+}\nu_{\ell}\gamma) < 3.0\times 10^{-6}$ at 90\% CL. This is a significantly more stringent limit than those obtained from previous measurements and demonstrates the great improvement obtained from the tagging techniques to be used in the Belle-II experiment.

\subsection{The decay $\boldsymbol {B^{+}\to \mu^{+}\nu_{\mu}\mu^{+} \mu^{-}}$}

The decays $B^{+}\to \mu^{+}\nu_{\mu}$ and ${B^{+}\to \ell^{+}\nu_{\ell}\gamma}$ are prohibitively difficult to reconstruct at a hadron collider due to the lack of a secondary vertex. Therefore, in order to access the same physics more complicated decays must be pursued. The decay $ {B^{+}\to \mu^{+}\nu_{\mu}\mu^{+} \mu^{-}}$, where two muons originate from either a virtual photon or a hadron such as the $\rho$ or $\omega$ mesons. The resulting experimental signature is much more appealing, with a well defined secondary vertex and three leptons in the final state.

The largest background, which originates from combinations of particles originating from different $B$ decays is suppressed by a requirement that the invariant mass of the two opposite sign muon combinations is below 1\,GeV. Another challenge for the measurement is to deal with the large amount of mis-identified background. This is controlled using a data sample in which the muon identification requirements have been reversed. The important feature of the analysis is to control the muon mis-identification rate in the presence of two other real muons. This is estimated by using a dedicated control sample of $B^{0}\to J/\psi K^{*0}$ decays.

\begin{figure}[tb]
  \centering
  \includegraphics[width=0.6\textwidth] {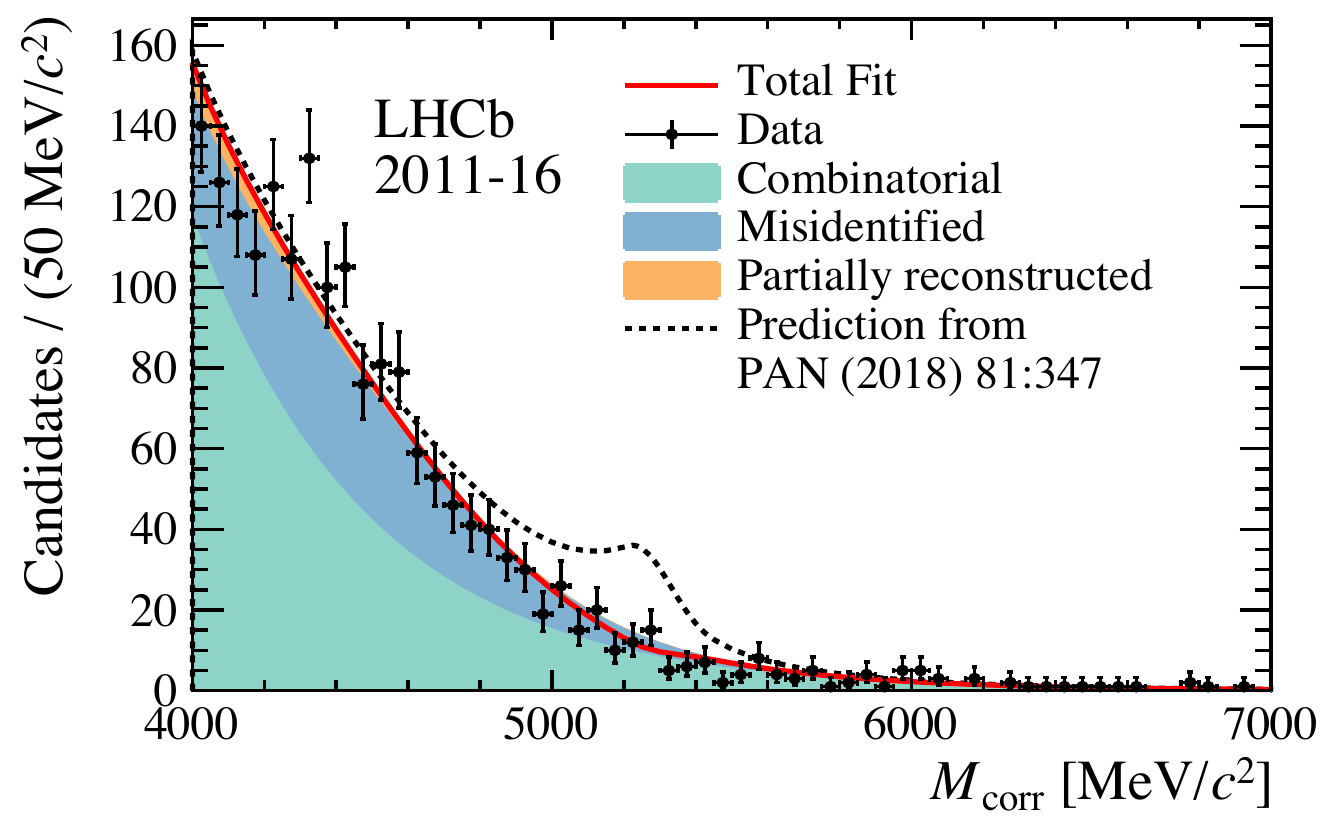}
  \caption{Projection of the fit to the missing mass, $M_{miss}$, for the determination of the $B\to\ell\nu_{\ell}\gamma$ signal yield.}
  \label{fig:BMuNuMuMu_fit}
\end{figure}

The signal yield is determined by fitting the so-called corrected mass, which is reconstructed by adding the momentum component perpendicular to the $B$ flight direction to the visible mass. The fit to the corrected mass distribution is shown in Fig.~\ref{fig:BMuNuMuMu_fit}, where a small negative signal yield is observed. This translates into an upper limit of $1.6\times 10^{-8}$ at 95\% CL. This limit is in tension with a recent theoretical prediction~\cite{Danilina:2018uzr}, which predicts a branching fraction around $1.3\times 10^{-7}$. The large dataset expected at the LHCb upgrade and further theoretical work will be needed to clarify this tension.

\section{Summary of the summary}

The study of semileptonic $B$ decays remains a very active field, and has a particularly strong interaction between the theoretical and experimental sides. This was reflected by the multitude of experimental and theoretical results presented at CKM 2018. It is clear that many new results are hotly anticipated from both the experimental and theoretical communities. These will no doubt make the next edition of the CKM conference in Melbourne as much of a success as the 2018 edition.

Finally, we would like to express our gratitude towards the excellent speakers of the WG II sessions, as well the flawless organisation of a highly enjoyable conference.

%\bibliographystyle{unsrt}
%\bibliography{references}
\end{document}